\documentclass[11pt,a4paper]{article}

\usepackage{a4wide}
\usepackage{graphicx}
\usepackage{hyperref}

\usepackage{mathtools}
\mathtoolsset{showonlyrefs}

\newcommand{\identifiers}[2]{\href{http://identifiers.org/#1/#2}{#1:#2}}
\newcommand{\doi}[1]{\identifiers{doi}{#1}}
\newcommand{\isbn}[1]{\identifiers{isbn}{#1}}
\newcommand{\biomodels}[1]{\identifiers{biomodels.db}{#1}}
\newcommand{\figref}[1]{Figure~\ref{figure:#1}}
\newcommand{\tabref}[1]{Table~\ref{table:#1}}

\newcommand{\diag}[1]{\, {\rm diag} \left( #1 \right)}
\newcommand{\rank}[1]{{\rm r} \left( #1 \right)}

\newcommand{\copasi}{COPASI }

\begin{document}

\setlength{\parindent}{0pt}
\addtolength{\parskip}{6pt}
\newlength{\figWidth} \setlength{\figWidth}{0.75\textwidth}

\author{Kieran Smallbone \\ [24pt]
Manchester Centre for Integrative Systems Biology\\
131 Princess Street, Manchester M1 7DN, UK\\
\href{mailto:kieran.smallbone@manchester.ac.uk}{\tt kieran.smallbone@manchester.ac.uk}
}

\title{Metabolic Control Analysis: Rereading Reder}

\date{}

\maketitle

\begin{abstract}

\noindent Metabolic control analysis is a biochemical formalism defined by Kacser and Burns in 1973, and given firm mathematical basis by Reder in 1988. The algorithm defined by Reder for calculating the control matrices is still used by software programs today, but is only valid for some biochemical models. We show that, with slight modification, the algorithm may be applied to all models.

\end{abstract}

\section*{Introduction}

Metabolic control analysis (MCA) is a biochemical formalism, defining how variables, such as fluxes and concentrations, depend on network parameters. It stems from the work of Henrik Kacser and James Burns in 1973~\cite{kacser73,kacser95} and, independently, Reinhart Heinrich and Tom Rapoport in 1974~\cite{heinrich74} (see David Fell's historical survey~\cite{fell92}). At the time, Kacser and Burns noted that there were two types of theory used to describe biochemical systems: static metabolic maps (detailing the overall structure) and enzymology (detailing the characteristics of the individual enzymes). The pair combined these two approaches, establishing a general theory of control of biochemical systems. Whole books have since been devoted to the topic~\cite{fell96,heinrich96}.

A general description of a biochemical system such as this may be given in ordinary differential equation format as

\begin{equation}
	\diag{c} \frac{d x}{d t} = N v \left( x, y, p \right), \qquad x(0) = x_0,
\end{equation}

where $c$ denotes compartment volumes, $x$ metabolite concentrations, $N$ the stoichiometric matrix, $v$ reaction rates, $y$ fixed metabolite concentrations and $p$ parameter values. 

The compartment volumes are required to match the units of $d x / d t$ (concentration per time) to those of $v$ (amount per time). However, this can also be incorporated in the stoichiometric matrix via the transformation $N \mapsto \diag{c}^{-1} N$. Moreover, for our purposes the fixed metabolite concentrations are parameters, in which case the system simplifies to

\begin{equation}
	\frac{d x}{d t} = N v \left( x, p \right), \qquad x(0) = x_0.
\end{equation}

Decomposing $v = e w$, where $e = 1$ denote the relative enzyme concentrations, is known as ``e-notation''~\cite{kacser95} and allows direct perturbation of reaction rates to examine system response.

We are interested in the steady states of the system and define 

\begin{equation}
	S = \lim_{t \to \infty} x, \qquad J = \lim_{t \to \infty} v = v(S, p).
\end{equation}

It is important to be explicit here: there is no reason to assume that non-linear systems such as these have a unique fixed-point. Rather, they could have many, or none. It is easy to fall into this trap; at the start of their paper, Kacser and Burns talk about ``the steady state'', though they later arrive at the more ambiguous ``a steady state''. If the limit above exists, then it is a well defined, unique fixed point. It should be noted that MCA has since been extended to cover non-stationary systems~\cite{ingalls03}.

Using MCA, we may analyse the effects of changes in enzyme concentration on system flux and concentration. We define the (scaled) concentration control coefficients $C^S$ and flux control coefficients $C^J$ as 

\begin{equation}
C^S = \left( \frac{e_j}{S_i} \frac{\partial S_i}{\partial e_j} \right)_{ij}, \qquad C^J = \left( \frac{e_j}{J_i} \frac{\partial J_i}{\partial e_j} \right)_{ij}.
\end{equation}

\section*{Example one}

\begin{figure}[!t]
	\centering
	\resizebox{\figWidth}{!}{
	\includegraphics{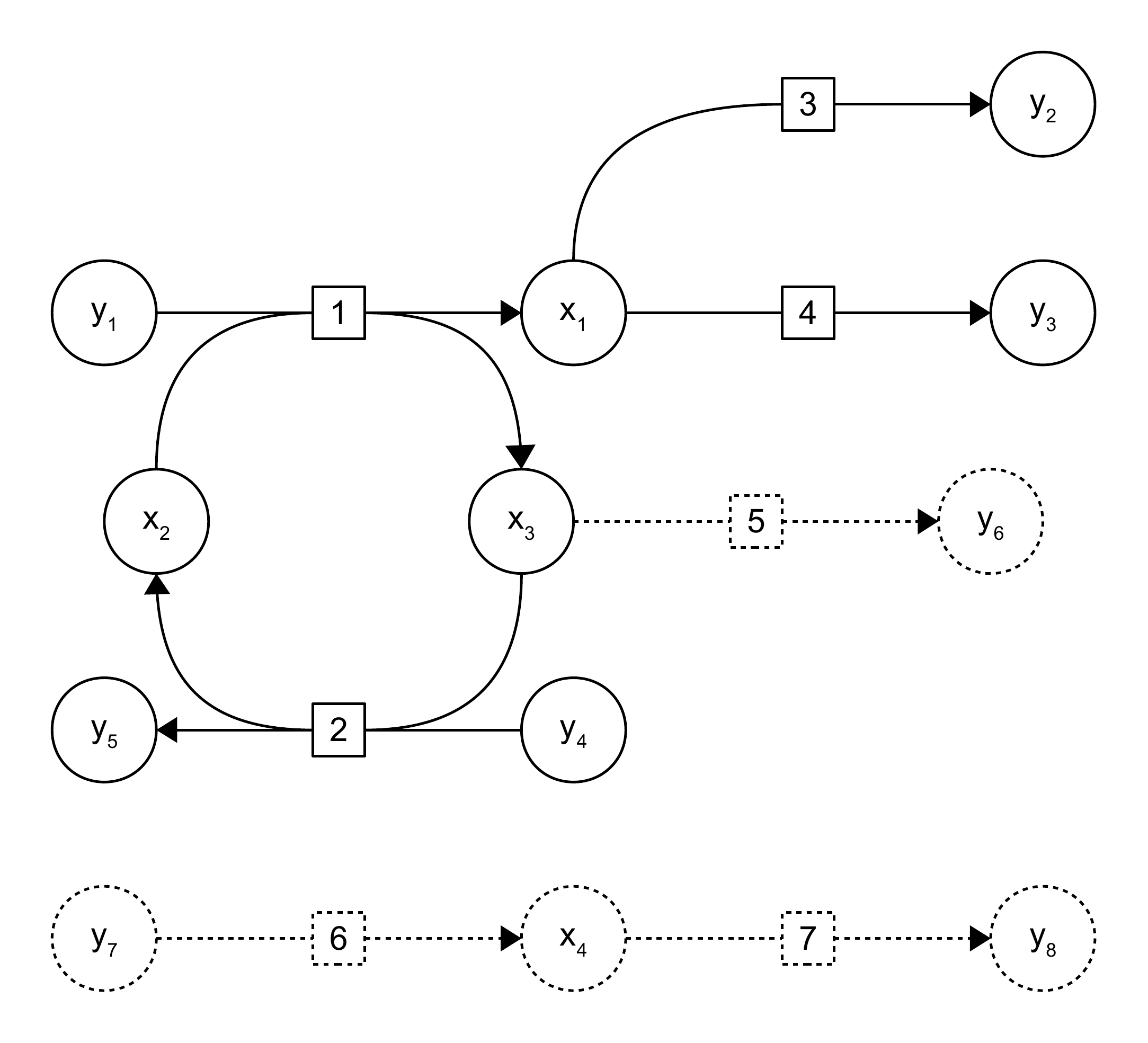}
}	
\caption{Diagrammatic representation of the models described in the text. The first example, containing only reactions 1 to 4,  is a small metabolic pathway with a branched flux and a moiety-conserved cycle~\cite{hofmeyr96}. The second example extends the first through inclusion of reaction 5, which has zero flux. The third example extends the first through inclusion of reactions 6 and 7, which have constant flux. Metabolites with prefix $x$ are the variables of the models, whilst those with prefix $y$ are fixed.}
\label{figure:1}
\end{figure}

We introduce an example, using the model of Jannie Hofmeyr and Athel Cornish-Bowden~\cite{hofmeyr96} (see \figref{1}) -- a small metabolic pathway with four reactions and three variables, including a branched flux and a moiety-conserved cycle.

Reaction kinetics in this example are defined by

\begin{eqnarray}
	v_1 &=& e_1 \frac{p_1 y_1 x_2 - x_1 x_3}{1 + y_1 + x_2 + x_1  + x_3 + y_1 x_2 + x_1 x_3}, 	\\
	v_2 &=& e_2 \frac{p_2 y_4 x_3 - y_5 x_2}{1 + x_3 + x_2 + y_4 + y_5 + x_3 y_4 + x_2 y_5}, 	\\
	v_3 &=& e_3 \frac{p_3 x_1 - y_2}{1 + x_1 + y_2},									\\
	v_4 &=& e_4 \frac{p_4 x_1 - y_3}{1 + x_1 + y_3}.
\end{eqnarray}

For the parameter values used, see \tabref{1}. The total concentration of the moiety-conserved species is taken to be $x_2 + x_3 = 5$. The steady state limits $S$ and $J$ are given in \tabref{2}.

We may go on to calculate the control coefficients for this system:

\begin{equation}
C^S = \left( 
\begin{array}{c c c c}
	0.241 	& 0.811 	& -0.877 	& -0.175 \\
	-1.121 	& 1.140 	& -0.015 	& -0.003 \\
	0.204 	& -0.207 	& 0.003 	& 0.001
\end{array}
\right),
\end{equation}
\begin{equation}
C^J = \left( 
\begin{array}{c c c c}
	0.228 	& 0.768 	& 0.003 	& 0.001 \\
	0.228 	& 0.768 	& 0.003 	& 0.001 \\
	0.228 	& 0.768 	& 0.170 	& -0.166 \\
	0.228 	& 0.768	& -0.830 	& 0.834 
\end{array}
\right).
\end{equation}

Notice that $C^S$ and $C^J$ satisfy what is known as the summation theorems, whereby 

\begin{equation}
	\sum_j C^S_{ij} = 0, \qquad \sum_j C^J_{ij} = 1,
\end{equation}	

which means that control is distributed over the system.

The summation theorems may be proved by appealing to Euler's homogeneous function theorem~\cite{euler1778,giersch88}:

\begin{quotation}

\noindent A function $f(x_1, \ldots, x_n)$ is called \emph{homogeneous} of degree $m$ in $x_i$ if $f(h x_1, \ldots, h x_n) = h^m f(x_1, \ldots, x_n)$ for all $h \neq 0$. The theorem states that if $f$ is homogeneous of degree $m$ then 

\begin{equation}
	x_1 \frac{\partial f}{\partial x_1} + \ldots + x_n \frac{\partial f}{\partial x_n} =m f.
\end{equation}

\noindent Conversely, every function that satisfies the above relationship is homogeneous of degree $m$ in $x_i$.
\end{quotation}

For our system, since the reaction rates depend linearly on the enzyme concentrations or activities, simultaneous transformation of these concentrations and of the time

\begin{equation}
	e^\ast =  h e, \qquad t^\ast = t / h,
\end{equation}

leads to a new equation system that coincides with the initial system after eliminating the superscript $^\ast$. Therefore, if the initial conditions are the same, metabolite concentrations of the transformed system at the moment $t/h$ will coincide with concentrations of the initial system at the moment $t$, whereas the fluxes will increase by factor $h$ (proportional to the new enzyme activities). The steady state concentrations will thus be unchanged, whilst the steady state reaction rates will increase by a factor $h$. Thus $S$ is homogeneous of degree 0 in $e$ and $J$ is homogeneous of degree 1 and, from the theorem

\begin{eqnarray}
	e_1 \frac{\partial S}{\partial e_1} + \ldots + e_n \frac{\partial S}{\partial e_n} 	& = & 0, \\
	e_1 \frac{\partial J}{\partial e_1} + \ldots + e_n \frac{\partial J}{\partial e_n} 	& = & J,
\end{eqnarray}

and hence the summation theorems hold.

\begin{table}[!t]
	\centering
	\begin{tabular}{c c}
		parameter & value \\
		\hline
		$p_1$ & 10 \\
		$p_2$ & 10 \\
		$p_3$ & 50 \\
		$p_4$ & 10 \\
		$p_5$ & 0 \\
		$p_6$ & 1 \\		
		$p_7$ & 1 \\		
		\hline
		$y_1$ & 10 \\
		$y_2$ & 0 \\
		$y_3$ & 0 \\
		$y_4$ & 1 \\
		$y_5$ & 1 \\
		$y_6$ & 0 \\
		$y_7$ & 1 \\
		$y_8$ & 0 \\
		\hline	
	\end{tabular}
\caption{Parameter values $p$ and fixed metabolite concentrations $y$ used in the models.}
\label{table:1}
\end{table}

\begin{table}[!t]
	\centering
	\begin{tabular}{c c}
		parameter & value \\
		\hline
		$S_1$ & 0.056 \\
		$S_2$ & 0.769 \\
		$S_3$ & 4.231 \\
		$S_4$ & 1 \\
		\hline
		$J_1$ & 3.196 \\
		$J_2$ & 3.196 \\
		$J_3$ & 2.663 \\
		$J_4$ & 0.533 \\
		$J_5$ & 0 \\
		$J_6$ & 1 \\
		$J_7$ & 1 \\
		\hline	
	\end{tabular}
\caption{Steady state and initial concentrations $S$ and fluxes $J$ of the model.}
\label{table:2}
\end{table}

\clearpage

\section*{Reder algorithm}

In software applications such as \copasi~\cite{copasi}, the control matrices $C^S$ and $C^J$ are not calculated through perturbation in $e$ and calculation of a new steady state; rather the algorithm outlined by Christine Reder in 1988 is used~\cite{reder88}, which allows calculation of the control matrices directly from the elasticities using simple matrix algebra. This algorithm is, however, only valid under certain circumstances. We step through it below to determine its validity.

We assume that $x_0 = S$ is a steady state, perturb about this state $x = x_0 + \tilde{x}$ and linearise.

\begin{equation}
	\frac{d x}{d t} 
	= \frac{d \tilde{x}}{d t} 
	= N v(x_0 + \tilde{x}) \approx N v(x_0) + N \varepsilon \tilde{x} = N \varepsilon \tilde{x} = \Delta \tilde{x},
\end{equation}

as $x_0$ is a steady state, where $\varepsilon$ denotes the linearisation of $v$ about $x_0$, known as the unscaled elasticity matrix in Reder's parlance, and $\Delta = N \varepsilon$ denotes the Jacobian of this dynamical system about $x_0$. 

In general, the rank $\rank{\Delta} = m_0 < m$, the number of metabolites, and the system will display moiety conservations -- certain metabolites can be expressed as linear combinations of other metabolites in the system~\cite{smallbone10}. Note that this number is not simply given by $\rank{N}$ as is generally, erroneously, suggested.

Let $R$ denote a full set of linearly independent rows of $\Delta$ and define the link matrix $L = \Delta \Delta_R^+$, where $\Delta_R^+ = \Delta_R^\prime \left(\Delta_R  \Delta_R^\prime \right)^{-1}$ denotes the Moore-Penrose pseudoinverse~\cite{penrose55}, which may be computed using QR factorisation~\cite{sauro04}. Thus $\Delta = L \Delta_R$.

Note that

\begin{equation}
	\frac{d \tilde{x}}{d t} = \Delta \tilde{x} = L \Delta_R \tilde{x} = L \frac{d \tilde{x}_R}{d t},
\end{equation}

and so we find $\tilde{x} = L \tilde{x}_R + T$, where the constant $T=0$.

Having developed a linear approximation of the system around the steady state, and a relationship between the independent and dependent metabolite concentrations, we now add a small perturbation $\delta$ to $e_j$, and calculate the change in steady state. 

We have 

\begin{eqnarray}
	\frac{d \tilde{x}_R}{d t} 	& = & 		N_R \diag{1 + \delta_j} w(x_0 + \tilde{x}) \\
						& \approx & 	N_R \diag{1 + \delta_j} \left( w(x_0) + \varepsilon \tilde{x} \right) \\
						%& = & 		N_R \varepsilon \tilde{x} + N_R \diag{\delta_j} \left( w(x_0) + \varepsilon \tilde{x} \right) \\
						& \approx & 	N_R \varepsilon \tilde{x} + N_R \diag{\delta_j} w(x_0) \\
						& = &		N_R \varepsilon \tilde{x} + \delta N_R v_j(x_0) \\
						& = &		N_R \varepsilon L \tilde{x}_R + \delta N_R v_j(x_0),
\end{eqnarray}

where we introduce the notation $\delta_j$ to denote the vector with entry $\delta$ in position $j$ and 0 elsewhere, and similarly $v_j (x_0)$ to denote the original steady state flux $v(x_0)$ in position $j$ and 0 elsewhere. Thus at, steady state,

\begin{equation}
	S = L S_R = - \delta L \left( N_R \varepsilon L \right)^{-1} N_R v_j (x_0).
\end{equation}

Letting $\delta \to 0$, we find

\begin{equation}
	\frac{d S}{d e_j} = - L \left( N_R \varepsilon L \right)^{-1} N_R v_j (x_0),
\end{equation}

and so the control matrices are given by

\begin{eqnarray}
	C^S 	& = & 	- \diag{x_0}^{-1} L \left( N_R \varepsilon L \right)^{-1} N_R \diag{v(x_0)}, \\
	% typo in original version: missing \varepsilon
	% C^J 	& = & 	\diag{v(x_0)}^{-1} \left(I- L \left( N_R \varepsilon L \right)^{-1} N_R \right) \diag{v(x_0)}. \\
	% thanks to Stefan Hoops for pointing this out
	C^J 	& = & 	\diag{v(x_0)}^{-1} \left(I - \varepsilon L \left( N_R \varepsilon L \right)^{-1} N_R \right) \diag{v(x_0)}. \\
\end{eqnarray}

These formulae allow calculation of the control matrices in terms of the local elasticities $\varepsilon$, rather than through perturbation of each $e_j$ and simulation. At first glance, they seem identical to those given in Reder; however, they are subtly different in the way the rows $R$ are chosen. In Reder,  $R^\ast$ represents the linearly independent rows of the stoichiometric matrix $N$ and Reder goes on to say that her analyses are only valid if the matrix $\Delta_{R^\ast} = N_{R^\ast} \varepsilon$ is of full rank. By contrast, our method directly chooses $R$ such that $\Delta_R$ is of full rank. In the case $\rank{N} = \rank{\Delta}$ (such as example one),we find $R = R^\ast$ and the two methods give identical results.

\section*{Example two}

We create a second example where the two methods differ, extending the first model through the reaction

\begin{equation}
	v_5\colon x_3 \rightarrow y_6, \qquad v_5 = e_5 p_5 x_3,
\end{equation}

where $p_5 = 0$. Thus $v_5 \equiv 0$ and so addition of this reaction should have no effect whatsoever on the system, and thus on other entries of the control matrices. However, the coefficients are given in \copasi (version 4.10) as

\begin{equation}
C^S_{\rm Reder} =  \left( 
\begin{array}{c c c c c}
	-0.888 	& 1.775 	& -0.878 	& -0.176 	& 0 \\
	-2.078 	& 2.078 	& -0.017 	& -0.003 	& 0 \\
	0 		& 0 		& 0 		& 0 		& 0
\end{array}
\right),
\end{equation}

and the flux control coefficients

\begin{equation}
C^J_{\rm Reder} =  \left( 
\begin{array}{c c c c c}
	-0.338 	& 1.325 	& 0.002 	& 0.000 	& 0 \\
	0.284 	& 0.716 	& 0.002	& 0.000	& 0 \\
	-0.840	& 1.681	& 0.169	& -0.166	& 0 \\
	-0.840	& 1.681 	& -0.831 	& 0.833	& 0 \\
	\infty		& \infty 	& \infty	& \infty 	& \infty
\end{array}
\right).
\end{equation}

These matrices are wildly different from the originals. The problem here is that addition of $v_5$ seems to break the conservation relationship between $x_2$ and $x_3$, at least in terms of the stoichiometric matrix. In reality, the two metabolites are still conserved. Instead choosing the independent rows as above, we find

\begin{equation}
C^S = \left( 
\begin{array}{c c c c c}
	0.241 	& 0.811 	& -0.877 	& -0.175 & 0\\
	-1.121 	& 1.140 	& -0.015 	& -0.003 & 0 \\
	0.204 	& -0.207 	& 0.003 	& 0.001 & 0
\end{array}
\right),
\end{equation}

\begin{equation}
C^J =  \left( 
\begin{array}{c c c c c}
	0.228 	& 0.768 	& 0.003 	& 0.001 & 0\\
	0.228 	& 0.768 	& 0.003 	& 0.001 & 0\\
	0.228 	& 0.768 	& 0.170 	& -0.166 & 0\\
	0.228 	& 0.768	& -0.830 	& 0.834 & 0 \\
	0 & 0 & 0 & 0 & 1
\end{array}
\right).
\end{equation}

We see that, as expected, the upper-left portions of each matrix are unaffected by the addition of the null reaction $v_5$. Moreover, each matrix still satisfies the summation theorems.

\section*{Example three}

For a final example model, we add (to the original model) reactions

\begin{eqnarray}
	&&v_6\colon x_7 \rightarrow x_4, \qquad v_6 = e_6 p_6 y_7, \\
	&&v_7\colon x_4 \rightarrow y_9, \qquad v_7 = e_7 p_7.
\end{eqnarray}

These reactions have constant rate, but are disconnected and independent of the rest of the system. Here the result of \copasi is worse than before, with $C^S$ and $C^J$ undefined (containing only NaN entries). We can have some sympathy here, as any perturbation in $e_6$ or $e_7$ leads to no steady state for $x_4$. However, the other entries for the matrix should be unaffected. Choosing $R$ differently, as above, we find the more appealing solutions:

\begin{equation}
C^S = \left( 
\begin{array}{c c c c c c}
	0.241 	& 0.811 	& -0.877 	& -0.175 & 0 & 0\\
	-1.121 	& 1.140 	& -0.015 	& -0.003 & 0 & 0 \\
	0.204 	& -0.207 	& 0.003 	& 0.001 & 0 & 0
\end{array}
\right),
\end{equation}

\begin{equation}
C^J =  \left( 
\begin{array}{c c c c c c}
	0.228 	& 0.768 	& 0.003 	& 0.001 & 0 & 0\\
	0.228 	& 0.768 	& 0.003 	& 0.001 & 0 & 0\\
	0.228 	& 0.768 	& 0.170 	& -0.166 & 0 & 0\\
	0.228 	& 0.768	& -0.830 	& 0.834 & 0 & 0\\
	0 & 0 & 0 & 0 & 1 & 0 \\
	0 & 0 & 0 & 0 & 0 & 1 \\
\end{array}
\right).
\end{equation}

\section*{Summary}

The existing method for calculating metabolic control matrices, though widely used, is only valid for certain models. The key issue is that dynamics, as well as stoichiometry, must be taken into account when determining moiety conservation. However, this small modification will ensure that the algorithm may be applied in all cases. 

\paragraph{Supplementary material}

The models described above are available in SBML format~\cite{sbml} from the BioModels Database~\cite{biomodels}. Their accession numbers are:

\begin{itemize}
	\item model 1: \biomodels{MODEL1305030000}
	\item model 2: \biomodels{MODEL1305030001}
	\item model 3: \biomodels{MODEL1305030002}	
\end{itemize}

\paragraph{Acknowledgements}

I am grateful for the financial support of the EU FP7 (KBBE) grant 289434 ``\href{http://www.biopredyn.eu}{BioPreDyn}: New Bioinformatics Methods and Tools for Data-Driven Predictive Dynamic Modelling in Biotechnological Applications''. Thanks to Mark Muldoon for innumerable discussions, and to Pedro Mendes for access to his innumerable library. Thanks also to Stefan Hoops for spotting a typo in the original version.

\end{document}